# Unconditionally secured classical cryptography using quantum superposition and unitary transformation

Byoung S. Ham

Center for Photon Information Processing, School of Electrical Engineering and Computer Science,
Gwangju Institute of Science and Technology, Gwangju 61005, South Korea

**Abstract**
Over decades quantum cryptography has been intensively studied for unconditionally secured data transmission in a quantum regime. Due to the quantum loopholes caused by imperfect single photon detectors and/or lossy quantum channels, however, the quantum cryptography is practically inefficient and even vulnerable to eavesdropping. Here, a method of unconditionally secured key distribution potentially compatible with current fiber-optic communications networks is proposed in a classical regime for high-speed optical backbone networks. The unconditional security is due to the quantum superposition-caused measurement indistinguishability of a paired transmission channel and its unitary transformation resulting in deterministic randomness corresponding to the no-cloning theorem in a quantum regime.

**Introduction**

In (classical) cryptographic technologies, there are two major versions: One is symmetric key-based private cryptography, and another is asymmetric key-based public one [1]. The public cryptography is called RSA and has become prevalent now, where its security relies on non-polynomial computational complexity of prime number factorization. Thus, the classical cryptography has been focused on the developments of efficient encrypting algorithms requiring more computing recourses in crypto-analysis. This is why the RSA key size has been continuously increased over decades, and now it is as long as 2048 bits [1]. As Internet traffic rapidly increases recently, information security has gained much more attention to protect the data from potential eavesdropping. Although the security of classical (public) cryptography looks good to some extent, it is basically conditional (or breakable) and even vulnerable to a quantum computer [2].

Quantum key distribution (QKD) belongs to the symmetric key-based private cryptography, and its security relies on how to distribute the keys rather than how to generate them. QKD has gained its importance due to theoretically confirmed unconditional security by Heisenberg's uncertainty principle in quantum mechanics [3]. Specifically the unconditional security of QKD is based on no-cloning theorem [4], resulting from quantum superposition between paired conjugate (non-orthogonal) variables used for bases of a quantum key [5]. The unconditional security of QKD, however, is not guaranteed in practice due to the quantum loopholes based on imperfectness of a single photon detector [6-12] and/or a quantum channel [12]. The detection loophole with the channel loss affects all QKD protocols based on single photons [5-7], entangled photon pairs [8-11], and coherent continuous variables [12]. As a result, QKD is practically fragile to eavesdropping unless the quantum loophole is completely closed [13]. Thus, the unconditional security of QKD has become a practical matter, resulting in the unrealistically low key rate. For example, in a standard optical fiber whose loss is $10^{-2}$ per 100 km, the actual quantum bit rate (QBR) drops down to $10^{-4}$, resulting in kilo~Mega-bits per second (bps) depending on the single/entangled photon generation rate [10]. Besides, technical difficulties in single-photon or entangled photon-pair generations make current QKD highly impractical. Most of all, current QKD is not compatible with conventional (classical) networks mainly due to nonlinear effects violating the no-cloning theorem. As a result, the transmission distance in QKD through an optical fiber is severely limited unless quantum repeaters are implemented [14].

Historically one-time-pad (OTP) has been proposed for an ideal communication system satisfying unconditional security, where the key is equivalent or longer than the data in length and must be used only one time [15]. Any existing cryptographic technologies, thus, do not support OTP simply due to either the low key rate or conditional security, while the classical data traffic rate in current fiber-optic communications backbone networks is more than 10 Gbps per channel, and its transmission is unlimited. Here, a completely different concept of unconditionally secured cryptography is proposed in a classical regime to overcome the limitations in both classical and quantum cryptographies and to support OTP. The proposed cryptography is safe from all kind attacks and quantum computers because its security is based on perfect randomness and measurement immunity.

Unlike QKD, the unconditional security in the proposed cryptography is provided by quantum-superposed transmission channels such as in a typical Young's double-slit experiment. As is well known, the Young's double-slit experiment is satisfied by both coherence (wave nature) [16,17] and incoherence (particle nature) optics [18], and the double slit can be replaced by a beam splitter (BS) in a Mach-Zehnder interferometer (MZI). In this paper, we focus on the classical nature of light (coherence optics) rather than the quantum nature to satisfy its classicality in both fundamental physics and potential applications. Compared with non-orthogonal basis set of a single photon in QKD, the orthogonal basis set of bright coherent light in the proposed cryptography has technical advantages to fit coming information era in terms of speed and compatibility. The key concept of the Young's double-slit experiments is in the measurement indistinguishability satisfied by both coherence (classical) and incoherence (quantum) physics. In other words, the state of a light such as a phase and a polarization cannot be measured definitely in MZI channels due to quantum superposition, resulting in prefect randomness in a binary system. According to the Shannon's information theory, the prefect randomness is equivalent to no eavesdropping or unconditional security [19]. To prove the unconditional security of the proposed cryptography, we present, analyze and discuss the fundamental physics of how to generate and distribute a perfect randomness-based key in a measurement-immune condition. Reminding of that QKD is the only method satisfying the unconditional security in key distribution using quantum mechanics, it is counterintuitive to perform the same function in a classical manner. This is the quintessence of the present paper.

As a physical infrastructure of the proposed unconditionally secured classical cryptography, a MZI scheme is used for the real transmission lines to realize both randomness-based key generation and unconditionally safe distribution via quantum superposition and unitary transformation (discussed in Figs. 1~3). It should be noted that MZI itself has already been used for some QKD protocols for encoding (for a sender) and decoding (for a receiver) through single transmission line [20-22], but it has nothing to do with the proposed one relied on double transmission lines with classical light. In the case of single-core fibers comprising the MZI scheme, the phase stability between them has already been proved for a km distance range by using a common locking technique [23]. Locking delicate noisy environments caused by temperatures, vibrations, and air fluctuations has also been proved in a free space for a 4-km distance range [24]. Technically the MZI stability issue is now closed and can be applied for a much longer traveling distance.

To understand the fundamental physics of the proposed cryptography, firstly, we present eavesdropping randomness and transmission directionality in an ideal MZI scheme. Then, round-trip MZI physics is analyzed for unitary transformation for engenvalue controllability. The unitary transformation in the proposed protocol is fulfilled with non-canonical (orthogonal) phase bases to satisfy a classical regime. The round-trip MZI-physics is then discussed for deterministic randomness, in which the key generation is random to eavesdroppers but deterministic to both sender and receiver. The deterministic randomness equivalent to the no-cloning theorem in QKD in a technical point of view is achieved classically via random shuffling of the eigenvalues in the MZI unitary transformation. Finally, the classically unconditional key distribution protocol is presented and discussed for potential attacks and future fiber-optic applications. This classically achieved unconditional security (perfect randomness in eavesdropping) surpassing QKD and RSA has never been discussed before and thus opens a door to the future information era.

**Results**

The phase shifter Φ in a MZI scheme of Fig. 1(a) is for a random basis selection between two orthogonal phase bases 0 and π. For the MZI unitary transformation, universal quantum gate operations have already been presented in a phase shifter-coupled MZI in a quantum regime [25]. Compared with nonorthogonal bases in QKD resulting in randomness according to the Heisenberg's uncertainty principle, the orthogonal bases in the proposed classical cryptography play the same role of the randomness in a classical regime (discussed in Figs. 2 and 3). For coherence optics with bright light fields, the split lights $E_3$ and $E_4$ on the first BS are perfectly coherent regardless of the bandwidth, intensity fluctuation, and phase noise of $E_1$. For incoherence optics with single photons, intensity correlation (or 4th order interference) has been proved for photon anti-bunching of the particle nature in a quantum regime [26]. These two different roles of BS have been intensively discussed for complementarity in quantum mechanics, where both phenomena cannot be dealt with simultaneously. The present protocol is for coherence optics but not excludes the particle nature of incoherence optics, either.

The BS matrix, [BS], was firstly discussed in 1979 by Degiorgio [16] and generalized in 1980 by Zeilinger [17], where there exists a π/2 phase shift between the split lights, the transmitted ($E_3$) and the reflected ($E_4$) for $\varphi = 0$ (see Fig. 1):

$$[BS] = \frac{1}{\sqrt{2}}\begin{bmatrix} 1 & i \\ i & 1 \end{bmatrix}, \tag{1}$$

where, $E_3 = \frac{E_1}{\sqrt{2}}$ and $E_4 = \frac{iE_1}{\sqrt{2}}$. There is no way to measure the absolute phase of traveling lights in the MZI channels unless $E_1$ is known. In other words, the measurement randomness in MZI channels is self-sustained by physics. Here, any measurement in the MZI channels also violates the indistinguishability in quantum superposition regardless of coherence or incoherence optics. This means that the channel measurement itself causes a fringe shift in the output interference pattern between $E_5$ and $E_6$. The relative phase measurement without fringe shift may be technically possible in an ideal system [27], but useless in crypto-analysis without knowing the input light ($E_1$) due to 50% chance in success (randomness). This randomness in eavesdropping represents no information withdrawal [19]. The path superposition of MZI channels, thus, becomes the origin of the unconditional security of the proposed protocol for a classical regime.

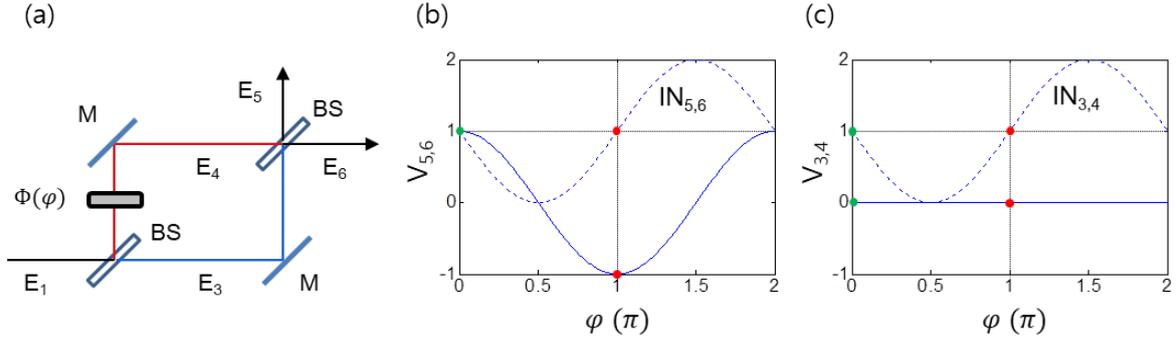

**Fig. 1. Deterministic randomness in MZI.** (a) MZI with a phase shifter Φ ($\varphi$): M, Mirror; BS, beam splitter. $E_i$ indicates light field in each region $i$. (b and c) Visibility $V_{i,j}$ (solid curve): $V_{i,j} = \frac{I_j - I_i}{I_j + I_i}$. $E_i$, coherent light pulse; $I_i$ is the intensity of $E_i$. $IN_{i,j}$ is the interference between $E_i$ and $E_j$ in the unit of $I_0$. The green and red dots refer to the basis $\varphi = \{0, \pi\}$.

In a typical MZI scheme of Fig. 1(a), each mirror generates the same phase shift in each path, resulting in perfect phase cancellation. The original light pulse $E_1$ generated by a commercial laser system hits on the first BS and split into two, $E_3$ and $E_4$. The split lights $E_3$ and $E_4$ are perfectly coherent each other in principle. This robust coherence of MZI even works for a single photon whose phase is random as an upper bound [23,27]. The random $\varphi$ –phase control for $E_3$ is provided by Bob using his phase shifter Φ, where the phase basis is binary and orthogonal: $\varphi = \{0, \pi\}$. In Fig. 1(a), the MZI matrix representation with a phase shifter Φ is denoted by:

$$[MZ]_\varphi = \frac{1}{2}\begin{bmatrix}(1 - e^{i\varphi}) & i(1 + e^{i\varphi}) \\ i(1 + e^{i\varphi}) & -(1 - e^{i\varphi})\end{bmatrix}, \quad (2)$$

where $[\Phi] = \begin{bmatrix}1 & 0 \\ 0 & e^{i\varphi}\end{bmatrix}$ and $[MZ]_\varphi = [BS][\Phi][BS]$. For $\varphi = 0$, the output lights at the second BS become unidirectional into $E_6$: $E_6 = iE_1$; $E_5 = 0$. The phase factor "$i$" in $E_6$ indicates a phase gain via MZI with respect to the input light $E_1$. For $\varphi = \pi$, the output light direction is switched into $E_5$: $E_5 = E_1$; $E_6 = 0$. As shown in Fig. 1(b) (see green and red dots in the solid curve), the output directionality in MZI is predetermined depending on the phase basis.

Allowing Eve to copy the traveling lights through MZI channels without altering the output interference fringe, the eavesdropping analysis in both visibility and interference between $E_3$ and $E_4$ proves the basic physics of measurement randomness: see Fig. 1(c). The orthogonal $\varphi$−values used for distinct visibility in Fig. 1(b), however, represent complete indistinguishability in the eavesdropping measurement. This $\varphi$−independent visibility in Fig. 1(c) is somewhat obvious owing to phase independency in measurements: $|E_3|^2 = |E_4|^2$. The measurement randomness is due to the fundamental physics of quantum superposition between two paths (phases) of MZI and corresponds to the no-cloning theorem in QKD. Even if Eve is highly sophisticated in eavesdropping with the same measurement tool as Alice's, Eve's success rate is 50% in average, resulting in perfect randomness like ideal coin tossing because the λ−limited phase change cannot be controlled in two independent systems (Bob-Alice & Bob-Eve; see Fig. 2) simultaneously: Further discussions are given in Discussion section.

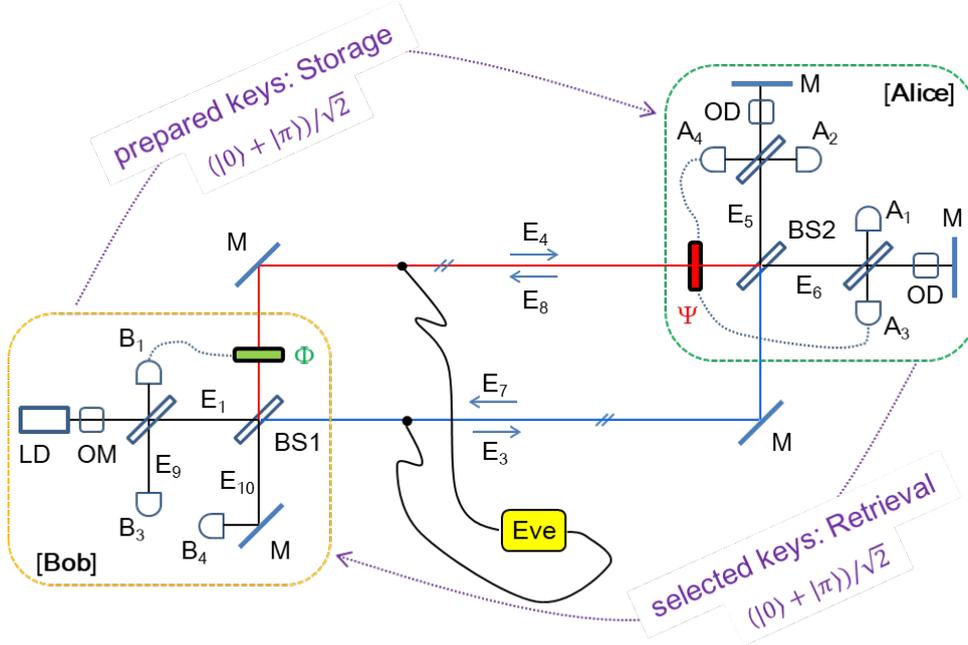

**Fig. 2. A schematic of PCD-MZI for OKD.** LD, Laser; OM, optical modulator; $A_i$, detector at Alice side; $B_i$, detectors at Bob's side; BS, 50/50 unpolarized beam splitter; M, mirror; Φ, Bob's phase controller; Ψ, Alice's phase controller; OD, optical delay; Eve, eavesdropper.

Figure 2 shows a schematic of the proposed protocol based on a round-trip MZI scheme, where the result of $[MZ]_\varphi^2$ is the identity matrix for $\varphi = \psi$ (see Section 1 of the Supplementary information):

$$[MZ]_\varphi^2 = (-e^{i\varphi})\begin{bmatrix} 1 & 0 \\ 0 & 1 \end{bmatrix}. \tag{3}$$

From equation (3), it is clear that a typical MZI system satisfies unitary transformation regardless of $\varphi$ if $\varphi = \psi$. The physical meaning of the identity matrix in equation (3) implies a time-reversible process as in a memory, where this phenomenon has been discussed for both quantum optics [28,29] and classical optics [30,31]. Here, the global phase in equation (3) has nothing to do with a measurement value or unitary transformation.

In the round-trip MZI configuration of Fig. 2, the phase shifter Ψ (Φ) is supposed to be invisible to the outbound (inbound) lights $E_5$ and $E_6$ ($E_9$ and $E_{10}$). For the key distribution, firstly, Bob prepares a key for Alice via random choosing of the phase basis $\varphi \in \{0, \pi\}$ and sends it to Alice via MZI channels. According to the MZI theory discussed in equation (2) and Fig. 1, Alice at the output port surely knows what Bob's random choice was by measuring her visibility $V_A$ (=$V_{5,6}$): MZI directional determinacy. For example, if Alice detects $A_2$ click for $E_5$ ($V_{5,6} = -1$) as shown in Fig. 1(b) (see the red dot), she definitely knows what Bob prepared is $\varphi = \pi$ representing the key '1', unless network error occurs: see Table 1(a) in details.

**Table 1. Visibility measurement-based key distribution in PCD-MZI.** (a) Alice's visibility $V_A$, (b) Bob's visibility $V_B$, (c) key sharing via deterministic randomness: $V_A = V_{5,6}$; $V_B = V_{9,10}$; $V_{ij} = \frac{I_j - I_i}{I_j + I_i}$

(a)

| Alice \ Φ($\varphi$) | $A_1$ | $A_2$ | Visibility ($V_A$) | Key (Preparation) |
|---|---|---|---|---|
| 0 | On | Off | 1 | 0 |
| $\pi$ | Off | On | −1 | 1 |

(b)

| Φ($\varphi$) \ Ψ($\psi$) | 0 (key=0) | $\pi$ (key=1) |
|---|---|---|
| 0 (key=0) | $B_3$ ($V_B$=−1) | $B_4$ ($V_B$=+1) |
| $\pi$ (key=1) | $B_4$ ($V_B$=+1) | $B_3$ ($V_B$=−1) |

(c)

| Bob's choice ($\varphi$) | key prepared | Alice's choice ($\psi$) | $V_A$ | $V_B$ | key final |
|---|---|---|---|---|---|
| 0 | 0 | 0 | 1 | −1 | 0 |
| 0 | 0 | $\pi$ | 1 | 1 | NA |
| $\pi$ | 1 | 0 | −1 | 1 | NA |
| $\pi$ | 1 | $\pi$ | −1 | −1 | 1 |

For the reflected light of $E_5$ and $E_6$, Alice randomly selects her phase basis $\psi \in \{0, \pi\}$ for her phase shifter Ψ and sends them back to Bob via the same MZI channels. The $\psi$-set inbound light $E_8$ together with $E_7$ is now going back through the same MZI, resulting in the final output lights, $E_9$ and $E_{10}$ at Bob's side. The matrix [BH] for the return light of $E_9$ and $E_{10}$ in Fig. 2 is represented by:

$$[BH]_{\psi/\varphi} = [MZ]_\psi [MZ]_\varphi = \frac{1}{2}\begin{bmatrix} -(e^{i\varphi} + e^{i\psi}) & i(e^{i\varphi} - e^{i\psi}) \\ -i(e^{i\varphi} - e^{i\psi}) & -(e^{i\varphi} + e^{i\psi}) \end{bmatrix}, \tag{4}$$

where $\begin{bmatrix} E_9 \\ E_{10} \end{bmatrix} = [BH]_{\psi/\varphi} \begin{bmatrix} E_1 \\ 0 \end{bmatrix}$. From equation (4), all four possible [BH] matrices are obtained:

$$[BH]_{0/0} = (-1)\begin{bmatrix} 1 & 0 \\ 0 & 1 \end{bmatrix}, \tag{5-1}$$

$$[BH]_{\pi/\pi} = \begin{bmatrix} 1 & 0 \\ 0 & 1 \end{bmatrix}, \tag{5-2}$$

$$[BH]_{0/\pi} = -i\begin{bmatrix} 0 & 1 \\ -1 & 0 \end{bmatrix}, \tag{5-3}$$

$$[BH]_{\pi/0} = i\begin{bmatrix} 0 & 1 \\ -1 & 0 \end{bmatrix}, \tag{5-4}$$

where each of them satisfies either identity ($E_9$) or inversion ($E_{10}$) relation: see Table 1(b) in details. Thus, Bob also surely knows which phase basis was set by Alice by observing his detectors $B_3$ and $B_4$ for visibility $V_{9,10}$ ($=V_B$). Then, the key is set deterministically if and only if the identity matrix is satisfied ($\varphi = \psi$): see Table 1(c) in details. Unlike mandatory sifting in QKD, Bob and Alice do not need to communicate with their measurement results. Although the key setting is inner shared with 100% sureness in an errorless MZI system, it is perfectly random to an eavesdropper Eve due to the measurement randomness as discussed in Fig. 1: deterministic randomness. Here, the MZI randomness in Fig. 1 is of course not sufficient to classical cryptography due to the *memory-based attack* (see Discussion). To protect it from this classical attack an additional action such as QKD-like sifting or *network initialization* must be given (discussed later). The deterministic randomness in Fig. 2 offers a significant feature of unconditional security to a classical regime. The discarded keys ($\varphi \neq \psi$) are of course used for network monitoring of eavesdropping (discussed in Fig. 3).

As shown in Table 1(b), the identity matrix of equations (5-1) and (5-2) is achieved if Alice chooses the same basis as Bob does ($\varphi = \psi$), and it is maximally distinguished from the inversion case of $\varphi \neq \psi$. Even though the identical basis ($\varphi = \psi$) results in the same value of $V_B = -1$ (see the diagonal values), Bob surely knows what Alice's choice is because he has prepared the key with $\varphi$: see also Table 1(c). Table 1(c) summarizes the key distribution determinacy in the proposed cryptography.

Figure 3 shows numerical calculations for Fig. 2 using equation (4). For the identity matrix of equations (5-1) and (5-2) with $\varphi = \psi$, the visibility of $V_{9,10}(V_B) = -1$ confirms the deterministic key distribution as shown in Figs. 3(a) and 3(b) (see the green and red dots). For the inversion matrix of equations (5-3) and (5-4) with $\varphi \neq \psi$, the visibility of $V_{9,10} = +1$ also confirms network monitoring (see the open circles).

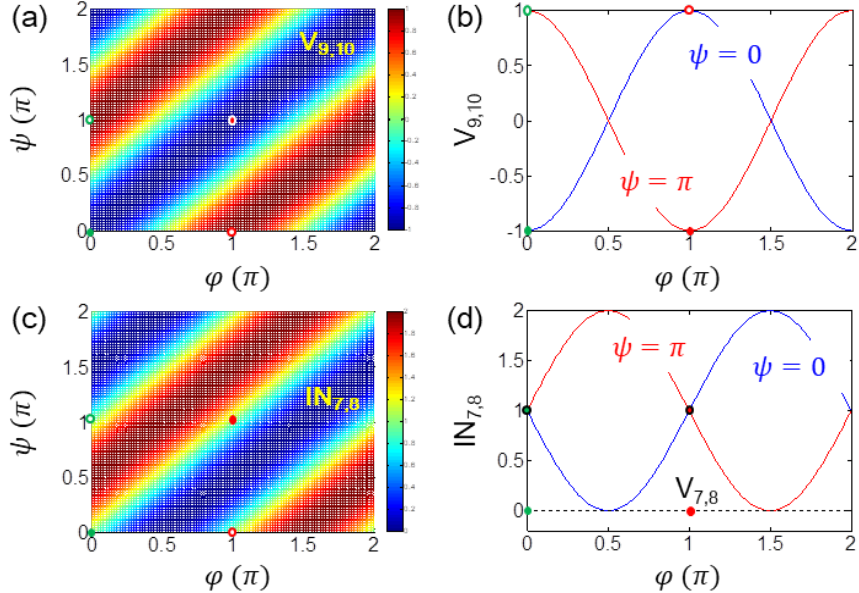

**Fig. 3. Numerical proofs for OKD in Fig. 2.** (a and b) Visibility $V_{9,10}$ for key distribution between Alice and Bob. The dashed and dotted curves are interference $I_{9,10}$ for $\varphi = 0$ and $\pi$, respectively. (c and d) The same value of interference $IN_{7,8}$ shows eavesdropping randomness. Green and Red dots indicate random keys set by Alice with $\psi \in \{0, \pi\}$ for $\varphi = \psi$. The open circles in (a) and (c) represent for discarded keys by Alice (see also open circles in (b)). Visibility $V_{ij} = \frac{I_j - I_i}{I_j + I_i}$: $I_i$ is the intensity of $E_i$.

As analyzed in Fig. 1(c) for eavesdropping randomness in MZI, the same analysis is performed for the return lights, $E_7$ and $E_8$ for indistinguishability, where the lights in both MZI paths have the same amplitude but different phase determined by $\varphi$ and $\psi$:

$$\begin{bmatrix} E_7 \\ E_8 \end{bmatrix} = \frac{1}{\sqrt{2}} \begin{bmatrix} -e^{i\varphi} & ie^{i\varphi} \\ ie^{i\psi} & -e^{i\psi} \end{bmatrix} \begin{bmatrix} E_1 \\ 0 \end{bmatrix}. \qquad (6)$$

As shown in Figs. 3(c) and 3(d), the matrix analysis of equation (6) for indistinguishability is numerically proved in both interference ($IN_{7,8}$) and visibility ($V_{7,8}$) (see the same value for different bases). Recalling the indistinguishability in the MZI path measurements in Fig. 1(c), Eve's measurement for the return lights ($E_7$ and $E_8$) reveals the same randomness: Eve never knows what basis is chosen by Alice as well as Bob due to the random basis selections as well as measurement indistinguishability in the superposed paths of MZI. This is the essence of the proposed cryptography using quantum superposition of MZI paths. The deterministic random key distribution process analyzed in Figs. 2 and 3 shows potential OTP applications owing to the compatibility with classical physics including duplication and amplification (see Discussion).

Except only for the keys denoted in green and red dots in Fig. 3(a), all others are considered as network errors caused by such as environmental noises and eavesdropping trials. Thus, Fig. 3(a) can be used as a bit error rate (BER) map. If Eve is successful for a safe measurement in both channels of MZI without the fringe shift, she can brutely scan her interferometer until a distinctive fringe patterns are observed. This brute force trial appears as a single curve in the BER map such as in Fig. 3(b). Even in this case, the probability of exact matching with the original one of Fig. 3(b) is 50% in average because there is no way to know exact MZI configuration due to independency of both systems. Thus, Eve's eavesdropping chance is random as in coin tossing. Here it should be noted that the random eavesdropping chance by Eve is, however, consistent to all bits, resulting in a room for a *memory-based attack* in classical crypto-analysis: see the *memory-based attack* in Discussion. By the way, the phase selection by both parties may be performed using a random number generator [32].

**Key distribution procedure**
The order (1~10) in Table 2 can be performed either individually or in a raw. The sifting process is necessary to avoid the *memory-based attack* (see Discussion). Assuming no network error or perfect tapping by Eve without affecting fringe shift, each bit rate in the key distribution procedure is as high as 50% with respect to Bob's key provision rate, which is more than Gbps according to current optoelectronic devices in fiber optic communications. As mentioned above, the outbound (inbound) light is invisible to the phase shifter $\Psi$ ($\Phi$). The key distribution procedure of the present cryptography is as follows (see Table 2):

[Sequence]
1. Bob randomly selects his phase basis $\varphi \in \{0, \pi\}$ to provide a $\varphi$-controlled coherent light pulse via the phase shifter $\Phi$ and sends it to Alice. Here, the $\varphi$-controlled light can be either individual or an N-bit chain for a batch job.
2. Bob converts his chosen $\varphi$ into a key set $\{x\}$ for a record: $x \in \{0,1\}$, where $x = 0$ if $\varphi = 0$ and $x = 1$ if $\varphi = \pi$.
3. Alice measures her detectors $A_1$ and $A_2$ for visibility $V_A$ to copy Bob's key $\{x\}$ in $\{y\}$ according to MZI directionality (see Table 1a): $y = 0$ if $V_A = 1$; $y = 1$ if $V_A = -1$; $y = V_A$ if $V_A \neq \pm 1$; $\{y\} = \{x\}$, except for $V_A \neq \pm 1$. Here, $V_A \neq \pm 1$ stands for an error due to eavesdropping or network problems: see the red number in Table 2.
4. Alice randomly selects her phase $\psi \in \{0, \pi\}$ to create a $\psi$-controlled light pulse via the phase shifter $\Psi$ and sends it back to Bob. Here, the $\psi$-phase control is performed on the reflected $\varphi$-controlled light pulse(s). This process is for the key setting, resulting in eavesdropping randomness as the inner-shared sifting process in addition to the MZI indistinguishability.
5. Alice converts her chosen $\psi$ into a key set $\{z\}$ for a record: $z \in \{0,1\}$, where $z = 0$ if $\psi = 0$ and $z = 1$ if $\psi = \pi$.

6. Alice sifts her prepared key in {z} into {a} by herself: a = y if y − z = 0; a=D if y − z ≠ 0. Here, D stands for discarded. This process is to avoid the *memory-based attack*.
7. Bob measures his detectors $B_3$ and $B_4$ for visibility $V_B$: w = x if $V_B = -1$; w=D if $V_B = 1$; w = $V_B$ if $V_B \neq \pm 1$. This step results in the copy of {a} into {w} (see Table 1(c)) except for error D (red). Here, $V_B \neq \pm 1$ stands for an error due to eavesdropping or network problems.
8. Bob sifts the copied key in {w} into {b} by himself: b = w if w − x = 0; b=D if w − x ≠ 0; {w} = {a}, except for $V_B \neq \pm 1$.
9. Alice and Bob announce their error bits (red) only for $V_A \neq \pm 1$ or $V_B \neq \pm 1$, and discard all corresponding bits in their keys {a} and {b}, respectively. They never announce their selected bases or visibilities. Alice and Bob finally share the same key {m}. In Table 2, the occurrence of network error (red D) is exaggerated for demonstration purpose, where the key rate of {m} is close to the half of the prepared one {x}.

**Table 2. A key distribution procedure for Fig. 2.**

| Party | Order / Sequence | | 1 | 2 | 3 | 4 | 5 | 6 | 7 | 8 | 9 | 10 | set |
|---|---|---|---|---|---|---|---|---|---|---|---|---|---|
| Alice | 3 | $V_A$ | 1 | −1 | −1 | 1 | −1 | 1 | 1 | −0.5* | 1 | −1 | |
| | | Copy x: y | 0 | 1 | 1 | 0 | 1 | 0 | 0 | −0.5 | 0 | 1 | {y} |
| | 4 | ψ | 0 | 0 | π | 0 | π | 0 | π | π | π | π | |
| | 5 | z(ψ) | 0 | 0 | 1 | 0 | 1 | 0 | 1 | 1 | 1 | 1 | {z} |
| | 6 | Sifting y: a | 0 | D | 1 | 0 | 1 | 0 | D | D | D | 1 | {a} |
| | 9 | Final key | **0** | **D** | **1** | **0** | **1** | **D** | **D** | **D** | **D** | **1** | {m} |
| Bob | 1 | φ | 0 | π | π | 0 | π | 0 | 0 | π | 0 | π | |
| | 2 | Prepared key: x(φ) | 0 | 1 | 1 | 0 | 1 | 0 | 0 | 1 | 0 | 1 | {x} |
| | 7 | $V_B$ | −1 | 1 | −1 | −1 | −1 | −0.8* | 1 | −1 | 1 | −1 | |
| | | Copy a: w | 0 | D | 1 | 0 | 1 | −0.8 | D | 1 | D | 1 | {w} |
| | 8 | Sifting w: b | 0 | D | 1 | 0 | 1 | D | D | 1 | D | 1 | {b} |
| | 9 | Final key | **0** | **D** | **1** | **0** | **1** | **D** | **D** | **D** | **D** | **1** | {m} |

*The numbers in red refer to network errors by disturbance or eavesdropping.

**By the sifting process the discarded bit D is shared between Alice and Bob automatically even without public announcement owing to the MZI determinacy. Only error bits denoted by red numbers are announced publically to discard the corresponding bit from the final key set {m}: see the red D.

***The discarded bit D can be represented by any big number, e.g., D =9 for a computing algorithm: $V_A = V_{5,6}$; $V_B = V_{9,10}$; $V_{ij} = \frac{I_j - I_i}{I_j + I_i}$.

**Discussion**

*Unconditional security*
The basic physics of unconditional security in the proposed classical cryptography lies in the quantum superposition between noncanonical (orthogonal) variables in MZI, corresponding to the no-cloning theorem in QKD, where the no-cloning theorem originates in Schrodinger's uncertainty principle with canonical (nonorthogonal) variables. Compared with the Heisenberg's uncertainty principle-caused no-cloning theorem in QKD, the unconditional security of the present cryptography belongs to classical physics of indistinguishability in MZI channel measurement. The measurement (eavesdropping)-caused fringe shift in MZI corresponds to the measurement-caused demolition of a quantum state in QKD.

When Alice's random phase choice is activated for the prepared keys by Bob, the unconditional security is fulfilled via round-trip MZI unitary transformation in a classical regime, where the random choice corresponds to post-measurement sifting in QKD. In other words, the eigenvalue (a basis for a provided key) provided by Bob is randomly chosen (a basis for a final key) by Alice for key setting, resulting in deterministic randomness as analyzed in Fig. 2 (see also Fig. S1 in the Supplementary information). The deterministic randomness means that the eigenvalue is deterministically inner shared between both parties but perfectly random to an eavesdropper owing to the controlled MZI unitary transformation via quantum superposition. Thus, the phase controlled round-trip MZI becomes the physical bedrock of the present unconditionally secured classical cryptography. The novelty of the present cryptography protocol is in the realization of the unconditional security in a classical regime with orthogonal (non-canonical) phase bases of bright light. As a result, the proposed cryptography is compatible with current fiber-optic communications networks, and thus, can support OTP with a high-speed (high-bit rate) optical key distribution at an extremely low error rate.

*Memory-based attack*
The *memory-based attack* is one of the major attacks in classical crypto-analysis. All classically encoded data can be intercepted and stored in a permanent memory device until a new technology such as a powerful computer or an efficient algorithm emerges. This is why there are several different encryption levels depending on the confidential level, e.g., in government documents. In the present cryptography, the *memory-based attack* can also be a powerful tool to a sophisticated eavesdropper, where the 50% chance in eavesdropping applies to all bits synchronously. Thus, Eve just unanimously flips all bits in the same key block $\{m'\}$ for correction if her guess is wrong. This is why the random basis selection is needed for sifting as shown in Table 2, resulting in bit-by-bit randomness.

Another way to protect the key from the *memory-based attack* is to use *network initialization* (discussed in Table 3) for each bit of the key. By either sifting or *network initialization*, the eavesdropping randomness in Fig. 2 is achieved. Thus, the eavesdropping chance exponentially decreases as the key length increases: For an n-bit-long key block, the eavesdropping chance η is $\eta = 2^{-n}$. If the key length is as short as 126-bit long (n=126), it takes thousand times longer than the universe age to decipher the key with even the most powerful supercomputer in the world (see Section 3 of the Supplementary information). Because there is no efficient algorithm for perfect random variables and the 126-bit long key can be easily and repeatedly (to some extent) generated by an even pseudo-random generator, it proves that the present protocol is unconditionally secured in a classical regime. By using personal computers and optoelectronic devices operating at GHz speed, the key distribution rate is independent of the transmission distance if a batch job is performed as shown in Table 2. Thus, the proposed protocol can be potentially applicable to a real-time key distribution system.

*Network Initialization*
For the deterministic randomness analyzed in Figs. 1~3, the *network initialization* between Alice and Bob is prerequisite to avoid the *memory-based attack* if there is no sifting as shown in Table 2. As a preparation step, Alice resets the MZI network with intentional phase turbulence to break the synchronized randomness in Eve's eavesdropping. To do this, Alice scans her phase shifter Ψ($\delta$) until she has maxima in visibility $V_A$ for the same test bits provided by Bob. The value of $V_A$, however, is not determined by the $\varphi$ phase basis because of $\varphi \neq \delta$. This fact also applies to Eve ($\delta$') in the same analogy: $\delta \neq \delta'$. Thus, the key sharing between Bob and Alice is not deterministic anymore. To solve this dilemma, i.e., to let only Alice know secretly and deterministically the $\varphi$-value set by Bob, the following *network initialization* procedure must be performed before the key procedure of Table 2 (see Table 3).

Table 3 is for *network initialization* preceded the key distribution procedure in Table 2: sequence 2~5. For this, firstly, Alice randomly resets the MZI system by arbitrarily adjusting a path length with an additional phase variable $\delta$ and scans it for her phase controller Ψ($\delta$) until she gets maxima in $V_A$: sequence 1. Then, Alice sends a cue to Bob. For this, Bob sends the same test bits encoded by $\varphi \in \{0, \pi\}$. Secondly, Bob randomly selects $\varphi \in \{0, \pi\}$ for the light pulse $E_4$ and sends it to Alice along with $E_3$ (see Fig. 2). Thirdly, Alice randomly sets her phase controller Ψ with either $\delta$ or $\delta+\pi$, measures $V_A$, and announces the result publically. Alice never announces her phase choice either for $\psi$ or $\delta$. Lastly, Bob measures his $V_B$ and publically announces whether Alice's measurement is correct or not. Then, Alice knows secretly and deterministically whether the $\delta$ is correct or wrong: Table 3 is for the case of non−π−phase shifted $\delta$. For the wrong case, Alice simply added a π phase to $\delta$ to fix it. The sequence 2~5 may be repeated until successful or to have a batch code for *network initialization* to provide the MZI network security indistinctly. As mentioned in Table 2, each *network initialization* must be performed for each bit if there is no sifting.

[Sequence]
0. Initially Alice resets the MZI network by disturbing the MZI with her phase controller $\Psi(\delta)$ and scans $\delta$ until she gets $V_A = \pm 1$ for the test bits provided by Bob. The $\delta$ is a phase variable added to her phase basis $\psi \in \{0, \pi\}$. Then, Alice gives a cue to Bob.
1. Bob randomly selects his phase basis $\varphi \in \{0, \pi\}$, encodes his light with $\varphi$, and sends it to Alice.
2. Alice measures $V_A$ and publically announces the result.
3. Alice resend the $\varphi$−set light after encoding it with $\delta + \psi$.
4. Bob measures $V_B$ and publically announces whether Alice's result is correct (O) or not (X).
5. Alice resets her phase basis $\psi \in \{0, \pi\}$ to either $\psi \in \{\delta, \pi + \delta\}$ or $\psi \in \{-\delta, \pi - \delta\}$ depending on the Bob's announcement: end of network initialization: Correctness
6. The sequence 2~5 may be repeated if not successful or to store additional initialization use with different $\delta$ values: Order (N=2~10).

**Table 3. Network initialization for Table 2.**

| Party | Sequence | Order (N) 1 | 2 | 3 | 4 | 5 | 6 | 7 | 8 | 9 | 10 |
|---|---|---|---|---|---|---|---|---|---|---|---|
| Alice | 2 $V_A$ | 1 | −1 | −1 | 1 | −1 | 1 | 1 | 1 | −1 | 1 |
| | 3 $\psi$ | $\delta$ | $\delta$ | $\delta+\pi$ | $\delta$ | $\delta+\pi$ | $\delta+\pi$ | $\delta$ | $\delta+\pi$ | $\delta$ | $\delta+\pi$ |
| | 5 Correctness | O | X | O | O | O | X | O | X | X | X |
| Bob | 1 $\varphi$ | 0 | π | π | 0 | π | 0 | 0 | 0 | π | 0 |
| | 4 $V_B$ | −1 | 1 | −1 | −1 | −1 | 1 | −1 | 1 | 1 | 1 |

*$V_A = V_{5,6}$; $V_B = V_{9,10}$
**Table 3 is for non−π−added $\delta$. For π−added $\delta$, see Section 4 of the Supplementary information.
*** "O" ("X") represents a correct (wrong) one.

Eve can also do the same job as Alice does with an arbitrary value of $\delta'$. In the same analogy Eve may get the same but unsynchronized fringe pattern due to $\delta \neq \delta'$. The chance of $\delta = \delta'$ is extremely law as shown in the BER map in Fig. 3(a). Here, the BER map resolution is determined by the detector sensitivity which is very high (>$10^4$ V/W at GHz) for commercially available photodiodes. Thus, the eavesdropping-immune MZI security is obtained by *network initialization*. To surprise, this MZI security is achieved by all classical means to satisfy the unconditionally secured cryptography. One might repute that Eve's intervention may cause a $V_A$ shift so that the initialization sequence could results an error. It could be true, but a consistent $V_A$ shift does not affect the unconditional security at all, otherwise, confirms Eve's existence. Thus, the *network initialization* can be used as authentication.

With the sifting in Table 2, the *network initialization* does not have to be repeated. The expected overall key distribution rate in Table 2, therefore, would be half of the usual data traffic rate. Without sifting, however, the *network initialization* must be performed for each bit to avoid the *memory-based attack*: see Section 4 of the Supplementary information. With the *network initialization* for each bit without sifting the key distribution speed for Table 2 (without sifting) may be slowed down.

*The man-in-the-middle attack*

The *man-in-the-middle attack* represents for 'intercept and resend,' where Eve behaves as Alice to Bob. As discussed in the *network initialization*, however, this attack does not work due to the inner-shard determinacy of MZI. The inner-shard determinacy is the intrinsic property of the MZI coherence as explained in Figs. 1~3. In other words, the MZI channel configuration between Bob and Alice cannot be exactly duplicated for Eve due to the system independency in coherence. Either single tapping or double tapping in Fig. 2, the phase synchronization between Alice and Eve cannot be achieved by any means. Thus, the *man-in-the-middle attack* should be failed for the proposed scheme.

*Applications*

Owing to strong demand in both wired and wireless communications, the information traffic in an optical fiber has increased three folds every two years over the last thirty years [33]. In optical fiber backbone networks, a traffic speed of 100 Gbps per (wavelength) channel has already been deployed for 80 channels in a dense wavelength division multiplexing system, resulting in a total capacity of 8 Tbps in a single-core optical fiber [34,35]. Thus, the capacity per fiber will reach its theoretical upper bound of 100 Tbps in a decade. Eventually a multicore fiber may replace current single-core fibers in the near future to overcome the channel capacity saturation [36]. In the multi-core fiber, a relative path-length drift caused by environmental noises such as vibrations and temperatures should be frozen due to spatial proximity between them in a few micron scales. Thus, the basic infrastructure of the double channels satisfying the MZI scheme for the present cryptography can be easily provided (see Fig. S3 of the Supplementary information).

For the applications, current 10~100 km spaced EDFA fiber-optic networks may be fit, where the MZI length becomes unlimited due to the coherence nature of light even with coherent amplifications at EDFA. This unlimited transmission distance is the $2^{nd}$ novelty of the present cryptography, where photon cloning by EDFA is basically phase-locked coherence process, resulting in only a fixed phase shift. The fixed phase shift in the cloning process can be dynamically adjusted in real time via visibility monitoring with laser locking techniques [23,24].

In conclusion, a coherence optics-based classical key distribution protocol was proposed, analyzed, and discussed to overcome the limitations in both classical and quantum cryptographies. The unconditional security of the proposed cryptography was obtained in a classical regime by using quantum superposition of MZI transmission channels and unitary transformation of MZI matrix. To prevent the system from typical cryptoanalysis such as memory-based attack and man-in-the-middle attack, sifting and network initialization protocols were presented and discussed for the unconditional key distribution. Here, the network initialization can also be adapted for authentication. By definition of coherence optics, the presented cryptography should be compatible with current fiber-optic communications networks at ~Gbps key rate. Thus, the proposed protocol has potential for one-time-pad cryptography which is the long lasting goal in human history for coming information era. Eventually, all-optical computers [37] may be combined with the present scheme for all-in-one secured information networks. The proposed cryptography is also applicable for wireless [38] or satellite [39] communications via MIMO [40] technologies (discussed elsewhere). Therefore, the round-trip unitary transformation based on MZI path superposition with non-canonical variables opens a door to new physics beyond QKD limited to a quantum regime.

**Acknowledgment:** The author acknowledges that the present work was supported by the ICT R&D program of MSIT/IITP (1711042435: Reliable crypto-system standards and core technology development for secure quantum key distribution network) and GRI grant funded by GIST in 2019.
**Additional Information**
Supplementary information is available in the online version of the paper. Reprints and permissions information is available online at www.nature.com/reprints.
**Competing interests**
The author declares no competing (both financial and non-financial) interests.
**Author contribution**
B.S.H. wrote the manuscript text and prepared all figures and tables. Correspondence and request of materials should be addressed to BSH (email: bham@gist.ac.kr)